\documentclass[amsmath,amssymb, aps,prb,reprint,showpacs,superscriptaddress,groupedaddress]{revtex4-1}

\usepackage{graphicx}
\usepackage{dcolumn}
\usepackage{bm,natbib}
\usepackage{threeparttable}
\usepackage{multirow}
\usepackage{adjustbox}

\begin{document}

\title{$\mathrm{Co_2Fe_{1-x}Cr_xSi}$ Heusler Alloys : A promising material for spintronics application}

\author{Deepika Rani$^1$, Jiban Kangsabanik$^1$, K. G. Suresh$^1$, N. Patra$^2$, D. Bhattacharyya$^2$, S. N.  Jha$^2$ and Aftab Alam$^1$}
\email{aftab@iitb.ac.in}
\affiliation{$^1$Department of Physics, Indian Institute of Technology Bombay, Powai, Mumbai 400076, Maharashtra, India\\ $^2$Atomic and Molecular Physics Division,
	Bhabha Atomic Research Centre, Mumbai 400085, Maharashtra, India
	}

\date{\today}

\begin{abstract}
In this article, we investigated the effect of Cr substitution in place of Fe on the structural, magnetic and transport properties of $\mathrm{Co_2FeSi}$ alloy. A comprehensive structural analysis is done using X-ray diffraction (XRD) and extended X-ray absorption fine structure (EXAFS) spectroscopy. Quaternary Heusler compounds $\mathrm{Co_2Fe_{1-x}Cr_xSi}$ with Cr content (x = 0.1, 0.3, 0.5) were found to crystallize in cubic structure. The synchrotron based EXAFS studies reveal that the anti-site disorder increases with the increase in Cr concentration. The saturation magnetization values in all the alloys are found to be less than those expected from the Slater-Pauling rule, which may be due to the some inherent disorder. A detailed resistivity analysis in the temperature range of 5-300 K is done, taking into account  different scattering mechanisms. The residual resistivity ratio is found to decrease with increasing Cr concentration. A disorder induced resistivity minimum due to weak localization effect is seen for x = 0.5. The resistivity measurements also indicate that the half-metallic character survives upto 100 K for x = 0.1, whereas the alloys with x= 0.3 and 0.5 show signature of half- metallic nature even at higher temperatures. First principles calculation done with a more robust exchange correlation functional (namely HSE-06) confirms the half metallicity in the entire concentration range. Theoretically simulated band gap and magnetic moments compliment the experimental findings and are compared wherever  possible. All these properties make $\mathrm{Co_2Fe_{1-x}Cr_xSi}$ a promising material for spintronics application.
\end{abstract}


\maketitle

\section{Introduction}
Many Heusler alloys have attracted much attention these days because of their half-metallic ferromagnetic (HMF) nature, high Curie temperatures and high spin polarization. For example, $\mathrm{Co_2FeSi}$  in the ordered $L2_1$ structure with a lattice parameter of 5.64 $\mathrm{\AA}$ has a moment of 6 $\mu_B$ and Curie temperature of 1100 K\cite{PhysRevB.72.184434} and therefore, $\mathrm{Co_2}$ - based alloys are one of the potential materials for spintronics applications. \cite{Graf2,doi:10.1063/1.4902831} Many of these alloys are predicted to be half-metallic on the basis of ab-initio calculations, some of which are verified experimentally. There is a possibility of designing new materials by exchanging the elements X, Y and Z or by substituting them by other elements to tune their properties. In the case of $\mathrm{Co_2FeSi}$, the Fermi level is near the bottom of the minority conduction band. Thus a small change can destroy the half-metallic character by shifting the Fermi energy to outside the minority gap. To shift the Fermi level to the middle of the gap and thus to obtain a stable structure, we have substituted Cr in place of Fe in $\mathrm{Co_2FeSi}$ alloy. The partial substitution for Fe by Cr may be seen as d-electron deficiency. The effect of Cr substitution on the spin polarization of $\mathrm{Co_2FeSi}$ was studied by Karthik et al. \cite{doi:10.1063/1.2769175} and they found that the spin polarization value increased with a small increase in the Cr concentration from 0.57 to 0.64. Later, Goripati et. al. studied the effect of Fe substitution with Cr on the giant magnetoresistance(MR) on $\mathrm{Co_2FeSi}$ based spin valves \cite{doi:10.1063/1.3549722} and found a large MR ratio at room temperature. Also, $\mathrm{Co_2Fe_{1-x}Cr_xSi}$ alloys were found to be half-metallic ferroamgnetic for  $x \leq 0.75$ by means of DFT simulations. \cite{GUEZLANE2016219, Gercsi-2007} In this paper we study the effect of Cr substitution for Fe on the structural, magnetic and transport properties of  $\mathrm{Co_2FeSi}$. In the case of Heusler alloys, an understanding of the chemical structure is very important in unravelling their spintronic and other physical properties. To elucidate the structure in detail, the site specific structural information was determined using Extended x-ray absorption fine structure (EXAFS) spectroscopy. The detailed structural analysis was done using x-ray powder diffraction and EXAFS spectroscopy. The alloys were found to crystallize in cubic structure ($L2_1$) and the EXAFS measurements reveal that the disorder increases with Cr concentration. The saturation magnetization were found to decrease with Cr doping. The experimentally observed moments were found to be less than those predicted by the  Slater-Pauling rule,\cite{Slat1,Paul1} which may be due to the antisite-disorder that increases with increasing Cr concentration. There exists no experimental confirmation of half-metallicity in these alloys. Thus, to further investigate the transport properties, the electrical resistivity measurements were performed in the temperature range of 5-300 K and the data was analysed by considering different scattering mechanisms. The resistivity measurements give an indirect indication that the alloys with x = 0.1 and x = 0.3 retain their half-metallic character only for $T < 100 K$, whereas the alloy with x = 0.5 shows the signature of half-metallic character even at higher temperatures. The residual resistivity ratio (RRR) is found to decrease with the increase in Cr concentration, which also reflects the increase in the disorder with Cr substitution.To gain further insight on the effect of Cr substitutions in $\mathrm{Co_2FeSi}$, we have simulated the spin-polarized electronic structure of $\mathrm{Co_2Fe_{1-x}Cr_xSi}$, (x = 0, 0.125, 0.25, 0.50 and 1) using HSE-06 exchange correlation functional which confirms the half-metallic nature for all the alloys, as observed experimentally.
\section{Experimental Details}
\subsection{Sample Synthesis}
The polycrystalline alloys $\mathrm{Co_2Fe_{1-x}Cr_xSi}$ were prepared by arc melting the stoichiometric amounts of the constituent elements (of at least 99.9\% purity) in water cooled copper hearth under high purity argon atmosphere. A Ti ingot was used as oxygen getter to further reduce the contamination. The ingot formed was flipped and melted several times for better homogeneity and the final weight loss was less than 1\%. 
\subsection{Characterization}
X-ray diffraction patterns were taken at room temperature using X”pert pro diffractometer with $\mathrm{Cu-K\alpha}$ radiation to study the crystal structure of the alloys. XRD analysis was done with the help of FullProf suite which uses the least square refinement between the experimental and the calculated intensities. In the Rietveld method, the weighted sum of squared difference between $y_{iobs}$ and $y_{ical}$ is minimized i.e., it tries to optimize the $\chi^2$ function given by:
\small
\begin{equation}
\chi^2= w_i\Sigma_i{(y_{iobs}-y_{ical})^2}
\end{equation}
\normalsize
where $w_i$ is the inverse of the variance associated with the $i^{th}$ observation i.e., $\sigma^2(y_{iobs})$ and $y_{iobs}$ and $y_{ical}$ are the observed and calculated scattering intensities for a diffraction angle $2\theta_i$.\cite{RR}

The synchrotron based EXAFS studies have been carried out at Co, Fe and Cr K-edges. The measurements have been carried out on the three samples having 10, 30 and 50 at.\% Cr doping concentrations at the energy scanning EXAFS beamline (BL-9) in transmission mode at the INDUS-2 Synchrotron Source (2.5 GeV, 100 mA) at Raja Ramanna Centre for Advanced Technology (RRCAT), Indore, India. \cite{:/content/aip/proceeding/aipcp/10.1063/1.4872706, 174265964931012032} This beamline operates in the energy range of 4 -25 keV. The beamline optics consists of a Rh/Pt coated collimating meridional cylindrical mirror and the collimated beam reflected by the mirror is monochromatized by a Si(111) ($2d=6.2709$ \AA) based double crystal monochromator (DCM). The second crystal of the DCM is a sagittal cylindrical crystal, which is used for horizontal focusing of the beam while another Rh/Pt coated bendable post mirror facing down is used for vertical focusing of the beam at the sample position. The EXAFS measurements at the Co (7709 eV) and Fe K-edges (7112 eV) were performed in transmission mode, while the measurement at Cr K-edge (5989 eV) was performed in the fluorescence mode. It should be noted that the measurement at Si K-edge (1839 eV) could not be performed since the above beamline works in the photon energy range of 4-25 keV.
In the transmission mode measurement, three ionization chambers (300 mm length each) have been used for data collection, one ionization chamber for measuring the incident flux ($I_0$), the second one for measuring transmitted flux ($I_t$) and the third chamber for measuring EXAFS spectrum of a reference metal foil for energy calibration.  Appropriate gas pressure and gas mixture have been chosen to achieve 10-20\% absorption in the first ionization chamber and 70-90\% absorption in the second ionization chamber to obtain higher signal to noise ratio. The rejection of the higher harmonics content in the X-ray beam was performed by detuning the second crystal of the DCM. The absorption coefficient $\mu$ is determined using the relation $\mu=ln(I_0/I_t)$, where the thickness of the absorber is unity. For the measurement, powder samples of appropriate weight, estimated to obtain a reasonable edge jump, have been taken and were mixed thoroughly with cellulose powder to obtain a total weight of 100 mg and homogeneous pellets of 15 mm diameter have been prepared using an electrically operated hydraulic press and were kept in a Teflon tape. For the EXAFS measurement in the fluorescence mode, the sample was placed at $45^0$ angle to the incident X-ray beam and the signal ($I_f$ ) was detected using a Si drift detector placed at $90^0$ to the incident beam. The first ionization chamber placed prior to the sample measures the incident beam ($I_0$ ). In this case, the X-ray absorption co-efficient of the sample is determined by the relation $\mu=I_f/I_0$ and the spectrum was obtained as a function of energy by scanning the monochromator over the specified range.\\

Magnetization isotherms at 5 K and 300 K were taken using a vibrating sample magnetometer (VSM) attached to the physical property measurement system (PPMS, Quantum Design) for fields up to 50 kOe. Electrical resistivity measurements were done using four probe method in PPMS.

\section{Computational Details}

Ab-initio calculations were performed using Density Functional Theory (DFT)\cite{kohn1965self} with Projector Augmented Wave (PAW) basis set\cite{blochl1994projector} as implemented in Vienna Ab-initio Simulation Package (VASP).\cite{kresse1996efficiency,kresse1999ultrasoft} To simulate the alloy Co$_2$Fe$_{1-x}$Cr$_x$Si, we have used a 4 atom primitive cell for x=0 and 1 (pure case), while a 2x2x2 supercell (32 atoms) for x= 0.125, 0.25 and 0.50.  Pseudopotential formalism with Perdew-Burke-Ernzerhof (PBE) exchange correlation functional\cite{perdew1996generalized} is used to do the structural optimization.  Co 3d$^{8}$4s$^{1}$, Fe 3d$^{7}$4s$^{1}$, Cr 3p$^{6}$3d$^{5}$4s$^{1}$, Si 3s$^{2}$3p$^{2}$  are considered as valence electrons. A plane wave cut-off of 450 eV along with a $\Gamma$-centered 4x4x4 K-mesh is used to sample the Brillouin zone while finding the equilibrium structures. All the structures were relaxed until the forces on each atom become less than 1 meV/\r{A}.  Using these optimized structures, we performed spin-polarized calculations using 400 eV plane wave cut-off and 8x8x8 $\Gamma$-centered k-mesh.Total Energies were converged upto $10^{-6}$ eV/\r{A}. To accurately predict the band gap related properties, we have used HSE-06 \cite{krukau2006influence} functional and simulated the previously relaxed structures. This time, we have taken 4x4x4 $\Gamma$-centered k-mesh. 

\section{Results and discussion}

\subsection{Structural Analysis}
Figure 1 shows the  x-ray diffraction patterns of the three compounds recorded at room temperature. The Rietveld refinement was done using the Fullprof Suite. The diffraction data confirm the cubic Heusler ($L2_1$) structure for all the compounds.   For $ x \geq 0.5 $, a minor additional phase started to appear. Due to the similar scattering amplitudes of the constituting elements, the (111) and (200) fcc superstructure reflections are not resolved for the three alloys. The lattice parameters as deduced from the refinement were found to be $5.65, 5.64$ and $5.64$ \AA\ for x $= 0.1, 0.3$ and  $0.5$ respectively. Thus, substituting for Fe by Cr leads to only a nominal change in the lattice parameter. The best fit of the observed intensities was obtained when Co, Fe, Cr and Si atoms were assigned the Wyckoff positions ($\frac{1}{4}, \frac{1}{4}, \frac{1}{4})$, (0,0,0,), (0,0,0) and ($\frac{1}{2},\frac{1}{2},\frac{1}{2}$) respectively. It should be noted that due to similar scattering coefficients of Co and Fe/Cr, the complete information of disorder cannot be obtained by X-ray diffraction. A deeper insight into the structure can be achieved by employing other methods such as EXAFS. Therefore, to further investigate the structure, local structure studies have been performed using EXAFS spectroscopy. 

\begin{figure}
\centering
\includegraphics[width=0.9\linewidth]{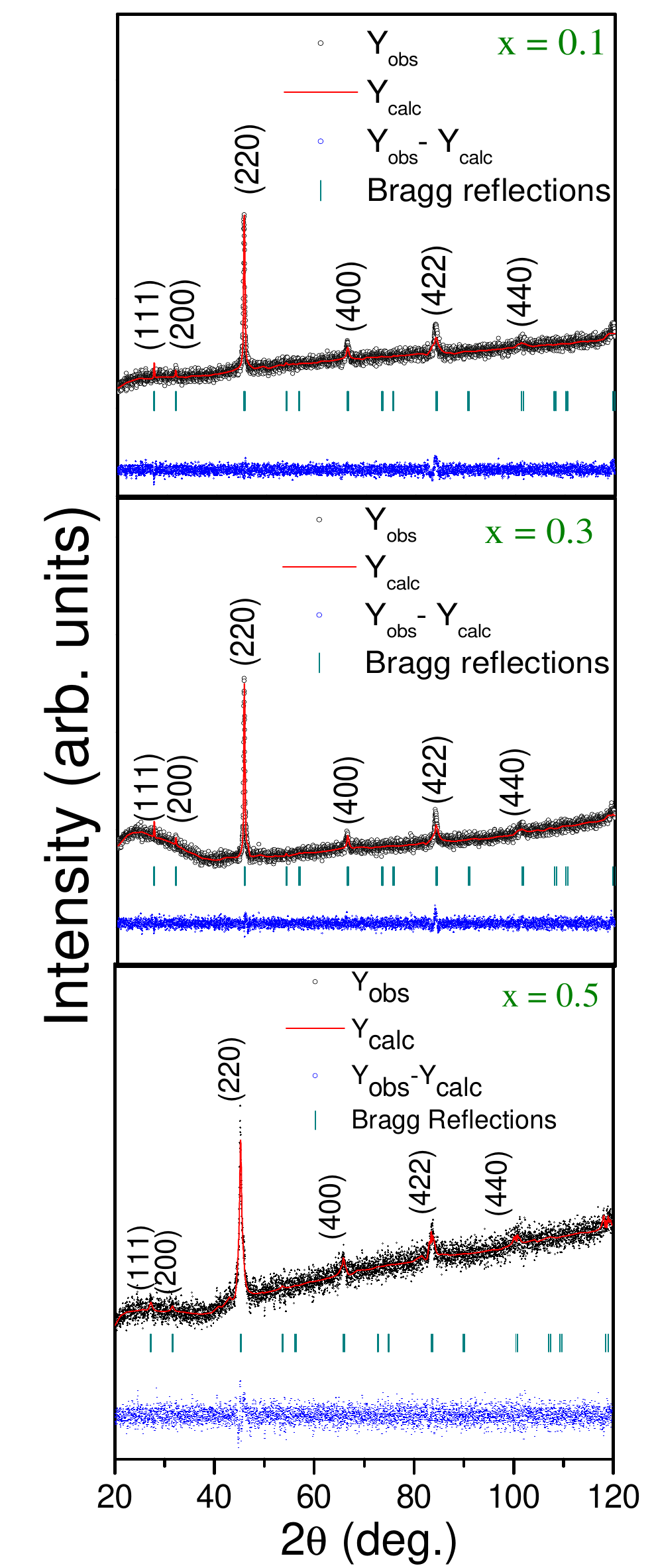}
\caption{Rietvield refined XRD patterns of $\mathrm{Co_2Fe_{1-x}Cr_xSi}$ alloys (x = 0.1, 0.3 and 0.5).}
\label{fig:CFCSXRDfinal}
\end{figure}

\subsection{EXAFS Analysis}
The analysis of the EXAFS data has been carried out following the standard procedure\cite{Dc,4book}using the IFEFFIT software package. \cite{5NEWVILLE1995154} This includes data reduction and Fourier transform to derive the $\chi(R)$ versus R plots from the absorption spectra, generation of the theoretical EXAFS spectra starting from an assumed crystallographic structure and finally fitting of the experimental $\chi(R)$ versus R data with the theoretical ones using the FEFF 6.0 code. \\

The structural parameters (i.e., space group, lattice parameter, Wyckoff positions of the atoms), which have been used to generate the Fourier transformed theoretical $\chi(R)$ versus R plots were obtained from the Rietveld refinement of the XRD data of the samples. The bond distances (R), co-ordination numbers (C.N) and structural disorder (Debye-Waller, $\sigma^2$) which give the mean-square fluctuations in the bond distances, have been used as fitting parameters and were varied independently for each of single scattering paths during the fitting process. The goodness of the fit in the above process is generally expressed by the $R_{factor}$,  which is defined as:

\small
\begin{equation}
R_{factor}=\sum {\frac {{[Im(\chi_{dat}(r_i)-\chi_{th}(r_i)]^2}+{{[Re(\chi_{dat}(r_i)-\chi_{th}(r_i)}]^2}}{{[Im(\chi_{dat}(r_i))]^2}+{[Re(\chi_{dat}(r_i))]^2}}}
\end{equation}
\normalsize

where, $\chi_{dat}$ and $\chi_{th}$ refer to the experimental and theoretical $\chi(r)$ values respectively and Re and Im refer to the real and imaginary parts of the respective quantities.
The background removed phase uncorrected $\chi(R)$ versus R plots were generated with the Fourier transform range in $k= 2.5-10.5$ $\mathrm{\AA}$  and the fitting was carried out in R space in the range $1.0- 5.0$ $\mathrm{\AA}$ using Hanning Window function. In addition to the single scattering paths, multiple scattering paths have also been considered for the sake of goodness of the fit.\\

\begin{figure}
	\centering
	\includegraphics[width=0.9\linewidth]{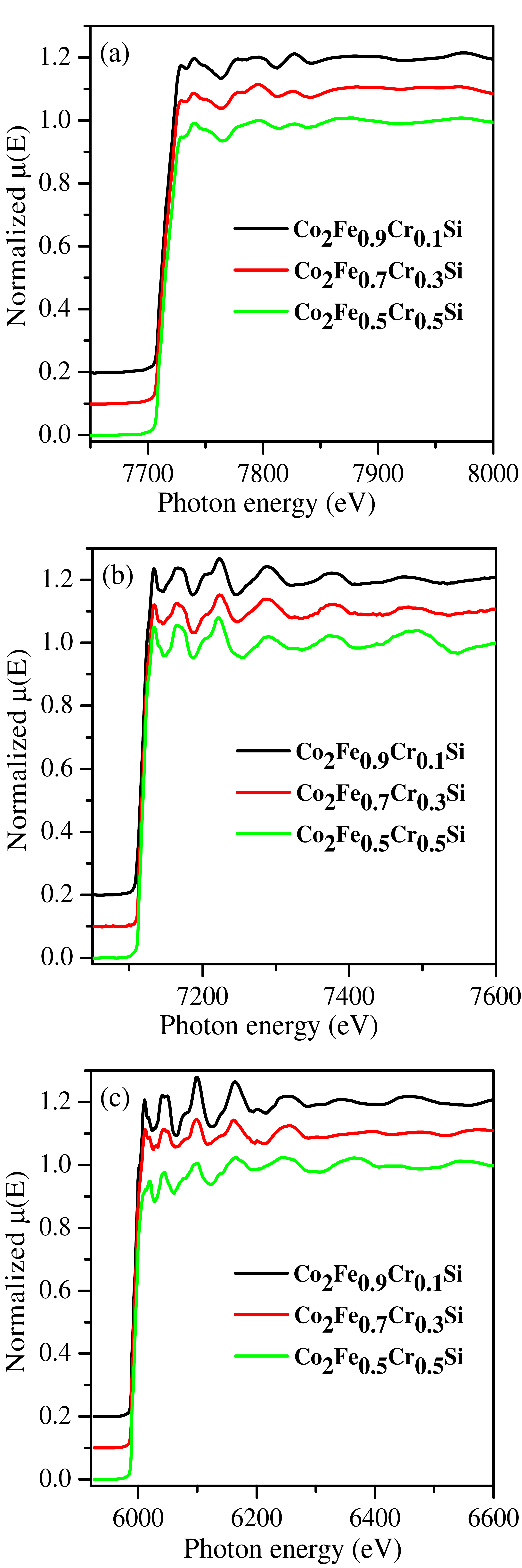}
	\caption{Normalized EXAFS spectra at (a) Co k-edge, (b) Fe k-edge and (c) Cr k-edge of $\mathrm{Co_2Fe_{1-x}Cr_xSi}$ Heusler alloys.}
	\label{fig:EXAFS2f}
\end{figure}
\begin{figure}
	\centering
	\includegraphics[width=0.9\linewidth]{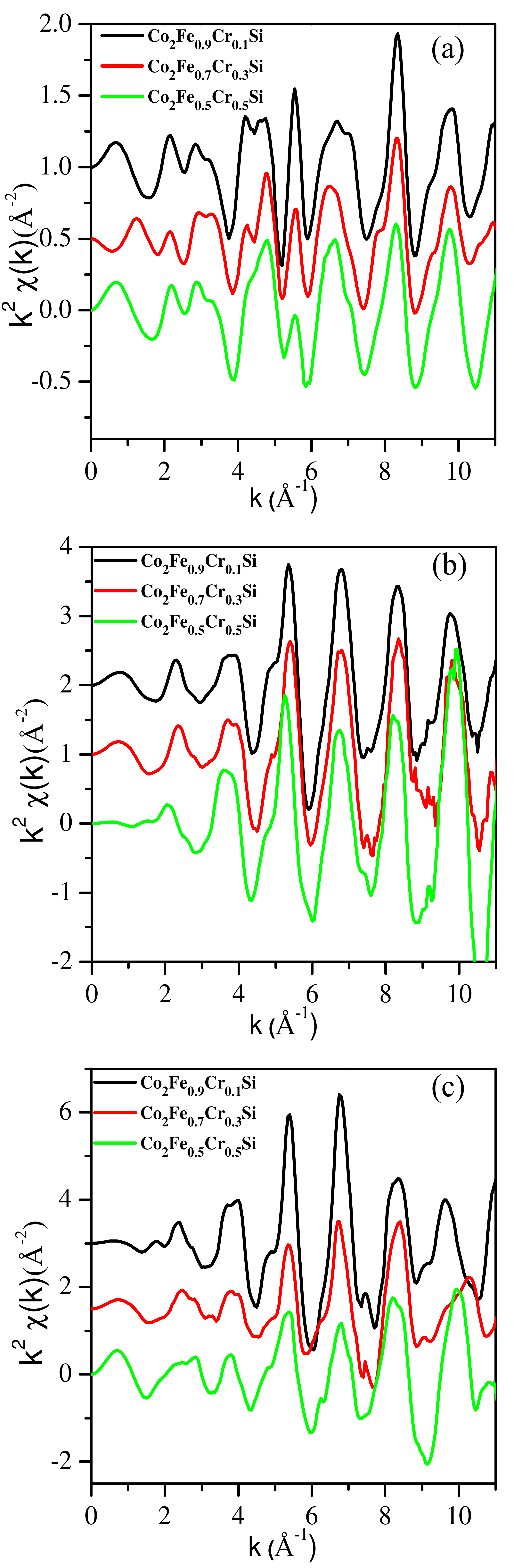}
	\caption{$\mathrm{k^2  \chi(K)}$ spectra at (a) Co K-edge, (b) Fe K-edge and (c) Cr k-edge of the $\mathrm{Co_2Fe_{1-x}Cr_xSi}$ Heusler alloys.}
	\label{fig:EXAFS1f}
\end{figure}
\begin{figure}
	\centering
	\includegraphics[width=0.9\linewidth]{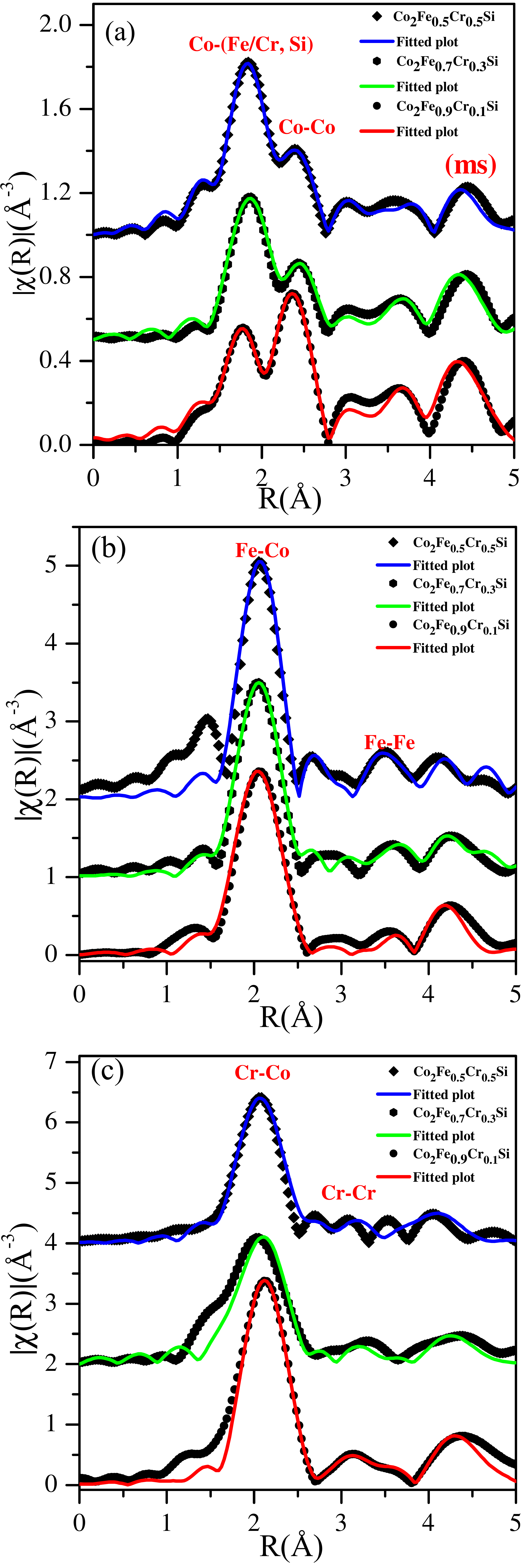}
	\caption{Measured $\chi(R)$ versus R along with the theoretically fitted plot of the $\mathrm{Co_2Fe_{1-x}Cr_xSi}$ Heusler alloys at (a) Co K-edge, (b) Fe K-edge and (c) Cr k-edge.}
	\label{fig:EXAFS3f}
\end{figure}

\begin{figure}
	\centering
	\includegraphics[width=0.9\linewidth]{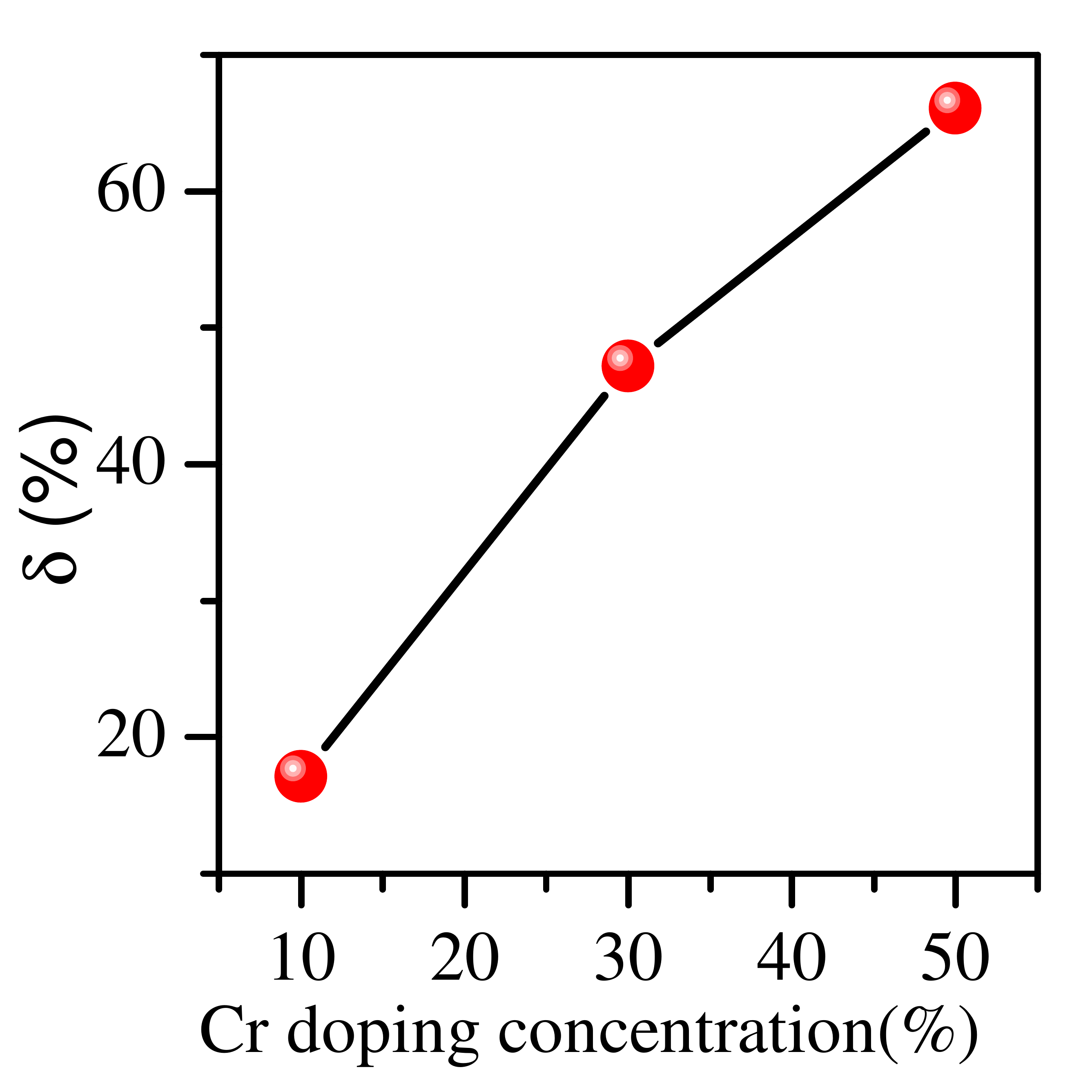}
	\caption{Variation of the disorder parameter ($\delta$) with the Cr doping concentration(x).}
	\label{fig:disorderEXAFS}
\end{figure}

\textbf {Co K-edge EXAFS}:

Figures \ref{fig:EXAFS2f}(a) and \ref{fig:EXAFS1f}(a) show the normalized EXAFS spectra ( $\mu$ versus E ) and the $k^2$ weighted $\chi(k)$ versus k plots at the Co K-edge for the $\mathrm{Co_2Fe_{1-x}Cr_xSi}$ Heusler alloys with different Cr concentrations, while Fig. 4(a) shows the corresponding $\chi(R)$ versus R plots. It can be seen from Fig. \ref{fig:EXAFS2f}(a) that the positions of the pre-edge and absorption edges are almost the same in all the three samples which manifests that oxidation state of Co in these samples remain the same irrespective of Cr concentration. $\chi(R)$ versus R plots of the alloys at the Co edge are found to be similar to those reported by others.  \cite{6PhysRevB.72.184434, 7:/content/aip/journal/apl/90/17/10.1063/1.2731314} It can be seen from figures 1-3 that the experimentally obtained spectra at Co K-edge are quite different from that obtained at Fe and Cr K-edges, both in k and R spaces.  This manifests a lowering in the crystallographic symmetry at Co site from the parent cubic structure. This difference may be due to the antisite disorder present in the samples, which has also been reported in earlier studies. \cite{6PhysRevB.72.184434, 7:/content/aip/journal/apl/90/17/10.1063/1.2731314, 80022-3727-40-6-S02}\\

For the fitting of the Co K-edge EXAFS data, a cubic $L2_1$ structure of four inter-penetrating sub lattices (space group Fm-3m \# 225) with a  lattice parameter of 5.65 $\mathrm{\AA}$  for 10\% Cr and 5.66 $\mathrm{\AA}$ for both 30\% and 50\% Cr doping have been used where Co atoms are situated at (1/4,1/4,1/4)  (4c) and (3/4,3/4,3/4 ) (4d) sites, Fe and Cr atoms at (0,0,0) (4a) site each and Si atom at (1/2,1/2,1/2) (4b) site with their corresponding occupancies which have been obtained from the XRD results. Balke et.al. have also obtained $L2_1$ structure for $\mathrm{Co_2FeSi}$ Heusler alloy prepared by arc melting. \cite{7:/content/aip/journal/apl/90/17/10.1063/1.2731314}  In order to generate the scattering paths for samples with different Cr concentrations, the Fe atoms in the FEFF input page have been replaced by the Cr atoms according to the occupancies. In the phase uncorrected Fourier transformed $\chi(R)$ versus R plots of Co K edge data, the first peak around 1.9 $\mathrm{\AA}$ arises due to Fe (Cr) first coordination shell.

Full Heusler alloys of the $\mathrm{Co_2FeSi (X_2YZ)}$ type are prone to antisite disorders.  The disorder may arise  due to three different kinds of binary mixing i.e. XY, XZ and YZ, the first  case giving rise to $DO_3$ type of disorder while the second and third  giving rise to B2 type of disorder.  It should be noted here that X-ray absorption spectroscopy is not suitable to identify the XY kind of disorder due to the similar X-ray scattering cross-sections of nearby elements in the periodic table \cite{9PhysRevB.65.184431} ($DO_3$ type disorder can hardly be detected by means of EXAFS techniques as the Co - Fe and Fe - Co bond distances are almost same in the first co-ordination cell.) However this technique is more sensitive to the YZ (B2) type of disorder or mixing between the atoms having largely different scattering factors. Hence, the disorder between Co and Fe/Cr atoms ($DO_3$ type disorder) has not been considered in the following EXAFS data analysis, but the disorder between Fe/Cr and Si atoms (B2 type) has been considered. \cite{6PhysRevB.72.184434}

To include the Fe(Cr)-Si antisite disorder of the first coordination shell in the fitting process,  inter-mixing of  Si atoms with  Fe and Cr atoms has been used by defining the amplitude of scattering paths for Fe and Cr atoms as ${{S_0}^2}*x$ and $S_0^2*y$ and the amplitude of scattering path for Si was defined as $S_0^2*(1-x-y)$, where ‘x’ and ‘y’ are the parameters  allowed to vary independently and $S_0^2$ is the electron back-scattering reduction factor taken to be invariant as 0.75 for all the paths. Thus throughout the fitting process the antisite disorder parameter was defined as $\delta = N_{Si} / (N_{Fe}+N_{Cr})$, N defines the co-ordination number of the respective atoms. This antisite disorder parameter δ gives the probability of occurrence of Si atoms at the Fe and Cr sites or vice versa by varying the Cr concentration.   

\begin{figure*}
	\centering
	\includegraphics[width=0.9\linewidth]{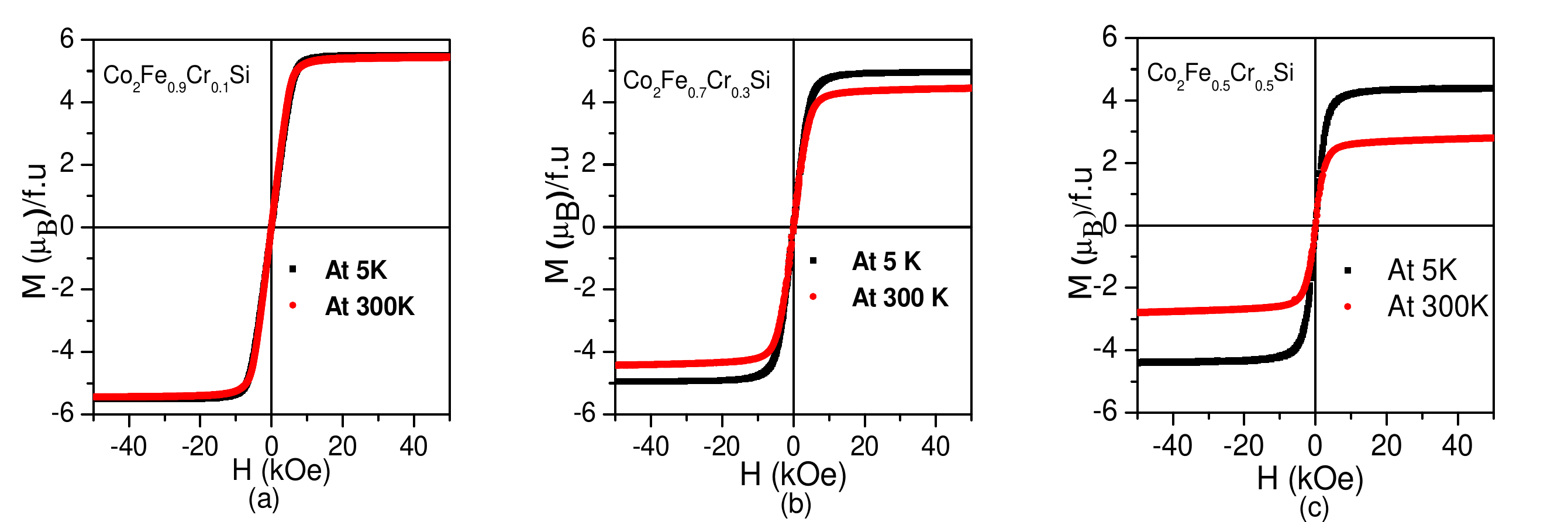}
	\caption{Isothermal magnetization curves at 5 K and 300 K for  $\mathrm{Co_2Fe_{1-x}Cr_xSi}$ alloys (a) x = 0.1, (b) x = 0.3 and (c) x = 0.5.}
	\label{fig:CFCSMH(3)}
\end{figure*}

\begin{figure}
	\centering
	\includegraphics[width=\linewidth]{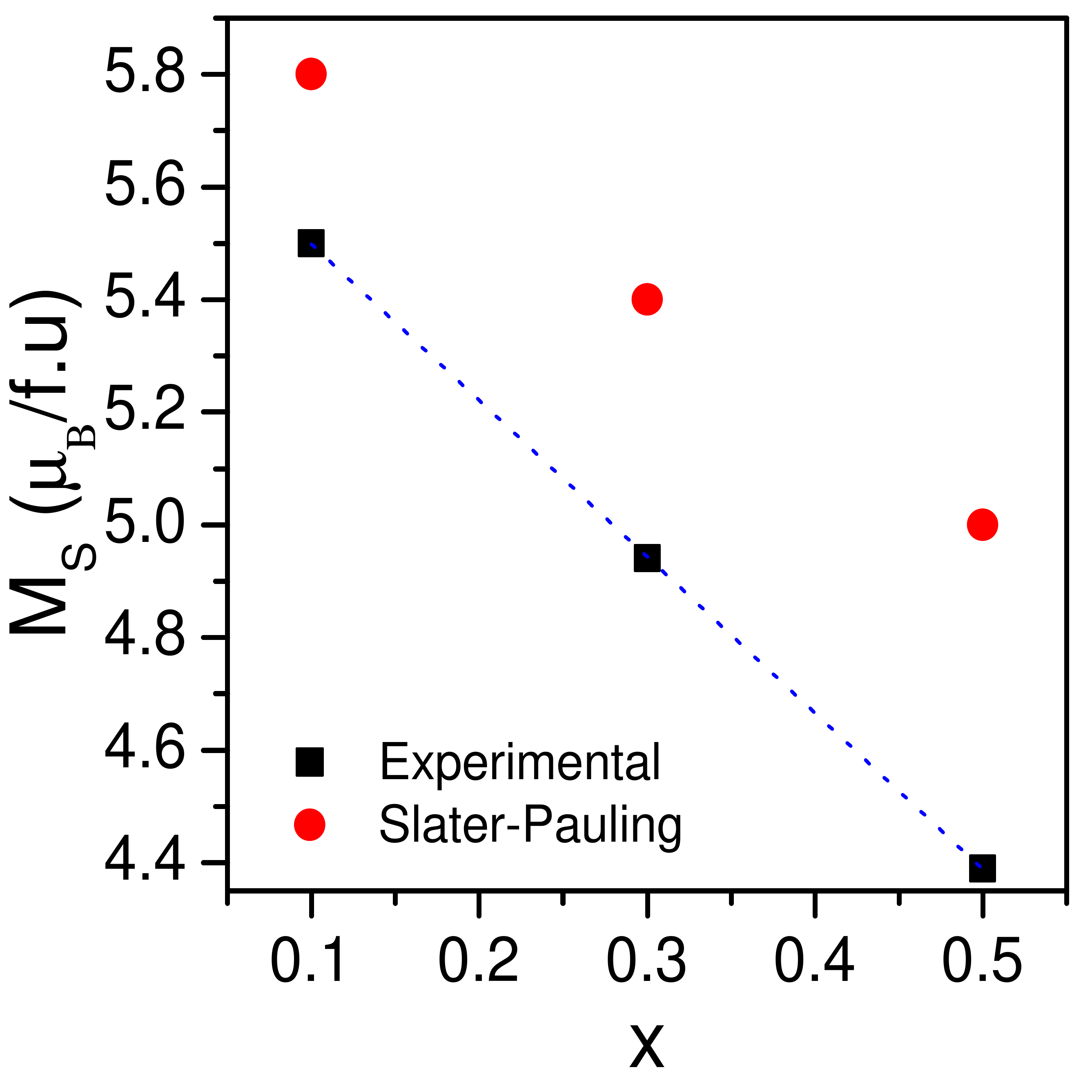}
	\caption{Deviation of saturation magnetization from S-P rule of $\mathrm{Co_2Fe_{1-x}Cr_xSi}$ alloys.}
	\label{fig:Spcmprsn}
\end{figure}

\begin{figure*}
	\centering
	\includegraphics[width=0.9\linewidth]{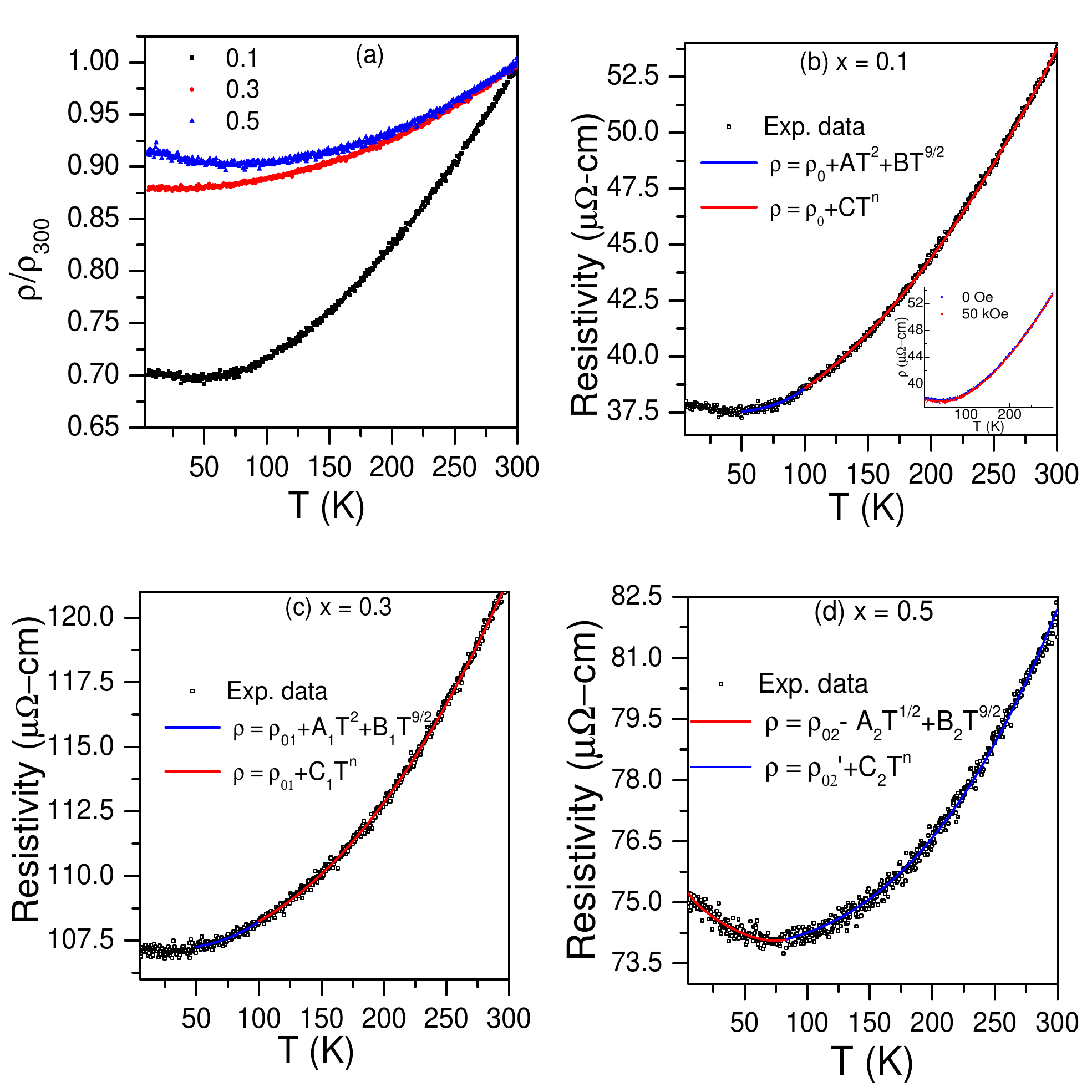}
	\caption{(a) Variation of normalised electrical resistivity for $\mathrm{Co_2Fe_{1-x}Cr_xSi}$ alloys and, measured and fitted temperature dependence of resistivity for $\mathrm{Co_2Fe_{1-x}Cr_xSi}$ alloys for (b) x = 0.1, (c) x = 0.3 and (c) x = 0.5. The inset in (b) shows $\rho$ vs T at 0 and 50 kOe for x = 0.1}
	\label{CFCSres}
\end{figure*}

\begin{figure}
	\centering
	\includegraphics[width=0.7\linewidth]{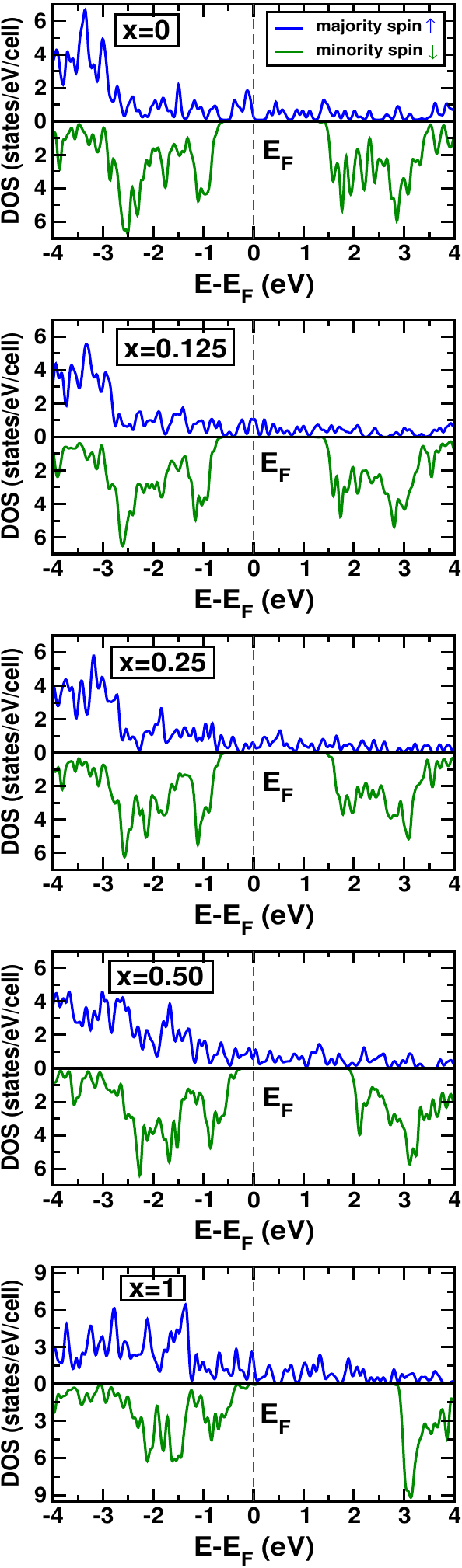}
	\caption{Spin polarized Density of States for Co$_2$Fe$_{1-x}$Cr$_x$Si, (x = 0, 0.125, 0.25, 0.50 and  1) using HSE-06 exchange correlation functional. No. of states in each figure is scaled with respect to 1 formula unit}
	\label{fig:fig9}
\end{figure}
	\begin{table*}
		\caption{\textbf{Co k-edge EXAFS results of \boldsymbol{$\mathrm{Co_2Fe_{1-x}Cr_xSi}$} Heusler alloys}}
		\centering
		\begin{tabular}{|c|c|c|c|c|}
			
			\hline \rule[-2ex]{0pt}{5.5ex}  \textbf{Scattering Paths}&  \textbf{Parameters}&  $\boldsymbol{\mathrm{Co_2Fe_{0.9}Cr_{0.1}Si}}$&  $\boldsymbol{\mathrm{Co_2Fe_{0.7}Cr_{0.3}Si}}$& $\boldsymbol{\mathrm{Co_2Fe_{0.5}Cr_{0.5}Si}}$ \\ 
			\hline \rule[-2ex]{0pt}{.5ex}  \textbf{Co-Fe1}&  C.N&  6.82&  2.18&  3.59\\ \cline{2-5}
			&$R(\mathrm{\AA})$ &2.41 &2.35 &2.41 \\ \cline{2-5}
			&$\sigma^2$ &0.019 &0.002 &0.013\\
			\hline \rule[-2ex]{0pt}{.5ex}  \textbf{Co-Si1}&  C.N&  1.17&  2.75&  3.04\\ \cline{2-5}
			&$R(\mathrm{\AA})$ &2.25 &2.46 &2.34 \\ \cline{2-5}
			&$\sigma^2$ &0.003 &0.002 &0.017\\
			\hline \rule[-2ex]{0pt}{.5ex}  \textbf{Co-Cr1}&  C.N&  &  3.07&  1.01\\ \cline{2-5}
			&$R(\mathrm{\AA})$ &  &2.46 &2.42 \\ \cline{2-5}
			&$\sigma^2$ & &0.001 &0.002\\
			\hline \rule[-2ex]{0pt}{.5ex}  \textbf{Co-Co1}$(2.82\mathrm{\AA})$&  C.N&  6.0&  6.0&  6.0\\ \cline{2-5}
			&$R(\mathrm{\AA})$ &2.70 &2.66 &2.74 \\ \cline{2-5}
			&$\sigma^2$ &0.010 &0.012 &0.023\\
			\hline \rule[-2ex]{0pt}{.5ex}  \textbf{Co-Co2}$(3.99\mathrm{\AA})$&  C.N&  12&  12&  12\\ \cline{2-5}
			&$R(\mathrm{\AA})$ &3.82 &4.22 &3.95 \\ \cline{2-5}
			&$\sigma^2$ &0.017 &0.018 &0.020\\
			\hline \rule[-2ex]{0pt}{.5ex}  \textbf{Co-Fe2}$(4.68\mathrm{\AA})$&  C.N&  11&  8&  6\\ \cline{2-5}
			&$R(\mathrm{\AA})$ &4.67 &4.69 &4.65 \\ \cline{2-5}
			&$\sigma^2$ &0.023 &0.005 &0.002\\
			\hline \rule[-2ex]{0pt}{.5ex}  \textbf{Co-Si2}&  C.N&  12&  12&  12\\ \cline{2-5}
			&$R(\mathrm{\AA})$ &4.42 &4.69 &4.52 \\ \cline{2-5}
			&$\sigma^2$ &0.025 &0.007 &0.008\\
			\hline \rule[-2ex]{0pt}{.5ex}  \textbf{Co-Cr2}&  C.N&  1&  4&  6\\ \cline{2-5}
			&$R(\mathrm{\AA})$ &4.61 &4.69 &4.65 \\ \cline{2-5}
			&$\sigma^2$ &0.028 &0.004 &0.001\\
			\hline \rule[-2ex]{0pt}{.5ex}  \textbf{Co-Co3}$(4.89\mathrm{\AA})$&  C.N&  8&  8&  8\\ \cline{2-5}
			&$R(\mathrm{\AA})$ &3.82 &4.22 &3.95 \\ \cline{2-5}
			&$\sigma^2$ &0.017 &0.018 &0.020\\
			\hline \rule[-2ex]{0pt}{.5ex}  \boldsymbol{$\mathrm{Co\rightarrow Co3 \rightarrow Fe1 \rightarrow Co}$}&  C.N&  8&  6&  4\\ \cline{2-5}
			&$R(\mathrm{\AA})$ &5.10 &4.69 &4.96 \\ \cline{2-5}
			&$\sigma^2$ &0.007 &0.003 &0.016\\
			\hline \rule[-2ex]{0pt}{.5ex}  \boldsymbol{$\mathrm{Co\rightarrow Co3 \rightarrow Si \rightarrow Co}$}&  C.N&  8&  8&  8\\ \cline{2-5}
			&$R(\mathrm{\AA})$ &4.90 &4.81 &4.76 \\ \cline{2-5}
			&$\sigma^2$ &0.004 &0.004 &0.018\\
			\hline 
		\end{tabular} \\
		
	\end{table*}

	\begin{table*}
		\caption{\textbf{Fe k-edge EXAFS results of \boldsymbol{$\mathrm{Co_2Fe_{1-x}Cr_xSi}$} Heusler alloys}}
		\centering
		\begin{tabular}{|c|c|c|c|c|}
			
			\hline \rule[-2ex]{0pt}{5.5ex}  \textbf{Scattering Paths}&  \textbf{Parameters}&  $\boldsymbol{\mathrm{Co_2Fe_{0.9}Cr_{0.1}Si}}$&  $\boldsymbol{\mathrm{Co_2Fe_{0.7}Cr_{0.3}Si}}$& $\boldsymbol{\mathrm{Co_2Fe_{0.5}Cr_{0.5}Si}}$ \\ 
			\hline \rule[-2ex]{0pt}{.5ex}  \textbf{Fe-Co1}$(2.44\mathrm{\AA})$&  C.N&  8&  8&  8\\ \cline{2-5}
			&$R(\mathrm{\AA})$ &2.42 &2.42 &2.42 \\ \cline{2-5}
			&$\sigma^2$ &0.008 &0.007 &0.005\\
			\hline \rule[-2ex]{0pt}{.5ex}  \textbf{Fe-Si1}$(2.82\mathrm{\AA})$&  C.N&  6&  6&  6\\ \cline{2-5}
			&$R(\mathrm{\AA})$ &2.61 &2.91 &2.89 \\ \cline{2-5}
			&$\sigma^2$ &0.018 &0.020 &0.008\\
			\hline \rule[-2ex]{0pt}{.5ex}  \textbf{Fe-Fe1}$(3.99\mathrm{\AA})$&  C.N&  11&  8&  6\\ \cline{2-5}
			&$R(\mathrm{\AA})$ &3.94 &4.01 &3.97 \\ \cline{2-5}
			&$\sigma^2$ &0.001 &0.017 &0.012\\
			\hline \rule[-2ex]{0pt}{.5ex}  \textbf{Fe-Cr1}&  C.N&  1&  4&  6\\ \cline{2-5}
			&$R(\mathrm{\AA})$ &3.94 &4.01 &3.97 \\ \cline{2-5}
			&$\sigma^2$ &0.001 &0.016 &0.012\\
			\hline \rule[-2ex]{0pt}{.5ex}  \textbf{Fe-Co2}$(4.68\mathrm{\AA})$&  C.N&  24&  24&  24\\ \cline{2-5}
			&$R(\mathrm{\AA})$ &4.66 &4.68 &4.66 \\ \cline{2-5}
			&$\sigma^2$ &0.002 &0.001 &0.001\\
			\hline \rule[-2ex]{0pt}{.5ex}  \textbf{Fe-Si2}$(4.89\mathrm{\AA})$&  C.N&  8&  8&  8\\ \cline{2-5}
			&$R(\mathrm{\AA})$ &4.70 &5.07 &5.05 \\ \cline{2-5}
			&$\sigma^2$ &0.002 &0.001 &0.001\\
			\hline
		\end{tabular}
	\end{table*}
	\begin{table*}
		\caption{\textbf{Cr k-edge EXAFS results of \boldsymbol{$\mathrm{Co_2Fe_{1-x}Cr_xSi}$} Heusler alloys}}
		\centering
		\begin{tabular}{|c|c|c|c|c|}
			
			\hline \rule[-2ex]{0pt}{5.5ex}  \textbf{Scattering Paths}&  \textbf{Parameters}&  $\boldsymbol{\mathrm{Co_2Fe_{0.9}Cr_{0.1}Si}}$&  $\boldsymbol{\mathrm{Co_2Fe_{0.7}Cr_{0.3}Si}}$& $\boldsymbol{\mathrm{Co_2Fe_{0.5}Cr_{0.5}Si}}$ \\ 
			\hline \rule[-2ex]{0pt}{.5ex}  \textbf{Cr-Co1}$(2.44\mathrm{\AA})$&  C.N&  8&  8&  8\\ \cline{2-5}
			&$R(\mathrm{\AA})$ &2.45 &2.42 &2.45 \\ \cline{2-5}
			&$\sigma^2$ &0.006 &0.005 &0.007\\
			\hline \rule[-2ex]{0pt}{.5ex}  \textbf{Cr-Si1}$(2.82\mathrm{\AA})$&  C.N&  6&  6&  6\\ \cline{2-5}
			&$R(\mathrm{\AA})$ &2.77 &2.63 &3.02 \\ \cline{2-5}
			&$\sigma^2$ &0.013 &0.011 &0.022\\
			\hline \rule[-2ex]{0pt}{.5ex}  \textbf{Cr-Fe1}&  C.N&  11&  8&  6\\ \cline{2-5}
			&$R(\mathrm{\AA})$ &3.93 &3.89 &4.14 \\ \cline{2-5}
			&$\sigma^2$ &0.018 &0.025 &0.024\\
			\hline \rule[-2ex]{0pt}{.5ex}  \textbf{Cr-Cr1}&  C.N&  1&  4&  6\\ \cline{2-5}
			&$R(\mathrm{\AA})$ &3.93 &3.89 &4.14 \\ \cline{2-5}
			&$\sigma^2$ &0.017 &0.024 &0.02\\
			\hline \rule[-2ex]{0pt}{.5ex}  \textbf{Cr-Co2}$(4.68\mathrm{\AA})$&  C.N&  24&  24&  24\\ \cline{2-5}
			&$R(\mathrm{\AA})$ &4.37 &4.62 &4.66 \\ \cline{2-5}
			&$\sigma^2$ &0.002 &0.012 &0.021\\
			\hline \rule[-2ex]{0pt}{.5ex}  \textbf{Cr-Si2}$(4.89\mathrm{\AA})$&  C.N&  8&  8&  8\\ \cline{2-5}
			&$R(\mathrm{\AA})$ &4.76 &4.59 &5.09 \\ \cline{2-5}
			&$\sigma^2$ &0.013 &0.004 &0.002\\
			
			\hline
		\end{tabular}
		
	\end{table*}

From FEFF simulation it has been found that the $2^{nd}$ peak at 2.4 $\mathrm{\AA}$ in the $\chi(R)$ versus R plots of the samples is due to Co-Co co-ordination shell at nominal distances of 2.83 $\mathrm{\AA}$, while the two peaks in the region of 3-4 $\mathrm{\AA}$  have contributions from the second Co-Co  coordination shells at 3.99 $\mathrm{\AA}$ along with the former one. The further Co-Fe(Cr)/Si coordination shell at 4.68 $\mathrm{\AA}$ contributes to the peak at 4.5 $\mathrm{\AA}$ along with two linear multiple scattering paths  with  back scattering angles of  $0^0$ and $180^0$.  During fitting, the co-ordination numbers of the higher scattering paths were kept constant according to their given crystallographic values and only the R and $\sigma^2$ values were varied.

The best fitted experimental results at Co K-edge have been summarized in  Table-1, which shows that Co-Fe bond lengths obtained in the first coordination shell is less than its theoretical value of 2.44 $\mathrm{\AA}$, possibly due to $L2_1-B2$ phase transition since Fe-Co bond length in B2 configuration is slightly shorter than that in the $L2_1$ configuration \cite{10:/content/aip/journal/apl/97/10/10.1063/1.3488805}. However the bond lengths in the higher co-ordination shells do not vary much. The bond lengths of the multiple scattering paths obtained are found to be close to the sum of the bond lengths of the constituting paths, which suggests that the multiple scattering paths are almost linear along the body diagonal direction. The variation in the value of   $\delta$ of the first coordination shell is shown in Fig.5, which shows that in the first coordination shell, significant antisite disorder is present and it increases almost linearly with increase in Cr concentration.

\textbf{Fe K-edge EXAFS}:
Figures 2(b) and 3(b) show the normalized EXAFS spectra ($\mu$ versus E) and the $k^2$ weighted $\chi(k)$ versus k  plots at the Fe K-edge of the $\mathrm{Co_2Fe_{1-x}Cr_xSi}$ alloys with varying Cr doping concentrations, while Fig. 4(b) shows the corresponding $\chi(R)$ versus R plots.  It can be seen that $\mu(E)$ versus E  and $\chi(R)$ versus R plots at the Fe K-edge are quite different from those of the Co K-edge, which is consistent with the results obtained by Wurmehl et. al.\cite{6PhysRevB.72.184434,80022-3727-40-6-S02}. This suggests that the local surrounding around the Fe atoms in the samples is different from that of Co atoms though both the sites have 8 atoms in the first co-ordination shell. Fig. 3(a) and 3(b) show that the ratio of intensities of the Fe-Fe path in the Fe site to that of the Co-Fe path in the Co site is almost 1:2. This confirms the tetragonal environment of the Co sites with respect to the Fe site, where each Fe or Cr atom is surrounded by 8 Co atoms while each Co atom is surrounded by 4 Fe(Cr) and 4 Si atoms. \cite{80022-3727-40-6-S02} In contrast to the Co edge, the Fe K-edge data was fitted without considering the antisite disorder since XRD measurements on these samples have not shown any anti-site disorder in the neighborhood of Fe sites. Similar model that has been used in the case of Co K-edge data has been used for the Fe K-edge data by placing Fe atom at the (0, 0, 0) position, without considering any multiple scattering paths.  The best fit results are summarized in the Table-2. It should be mentioned here that during the fitting of the Fe and Cr K edge data, the nominal values of the coordination numbers of various paths have been assumed to remain constant since there is no signature of anti-site disorder at Fe and Cr sites. In the phase uncorrected $\chi(R)$  versus R  plot, the first intense peak appears due to scattering from the first shell of 8 Co atoms at a nominal distance of 2.444 $\mathrm{\AA}$ and the other lower intense peaks at $\sim 2.8$ $\mathrm{\AA}$ and $\sim 3.5$ $\mathrm{\AA}$ arise due to the scattering from the coordination shells of 6 Si and 12 Fe/Cr atoms at the nominal distances of 2.823 $\mathrm{\AA}$ and 3.993 $\mathrm{\AA}$ respectively. Peaks at higher  values $> 4$ $\mathrm{\AA}$ have contributions from Fe-Co and Fe-Si shells at further distances. No mixing of the shells in the Fe K-edge was observed in comparison to the Co K-edge spectra. From  Table- 2 it  can be seen  that upon Cr doping the Fe-Co bond distances in both the first and higher co-ordination shells do not decrease much ($< 0.02$ $\mathrm{\AA}$) and remain almost the same having lower $σ^2$ value. This can be attributed to the sluggish nature of the Fe atoms which do not get much displaced and do not undergo the $L2_1$-B2 transition.

\textbf{Cr K-edge EXAFS}:

Figures 2(c) and 3(c) show the normalized EXAFS spectra ($\mu$ versus E)  and the $k^2$ weighted $\chi (k)$ versus k  plots at the Cr K-edge of the $\mathrm{Co_2Fe_{1-x}Cr_xSi}$ alloys with varying  Cr concentrations, while  figure 4(c) shows the corresponding $\chi(R)$ versus R plots. From the figure it can be seen that $\mu$ versus E spectra and $\chi(R)$ versus R plots at both Cr and Fe edges resemble each other very closely, which is due to the similar X-ray scattering factors and atomic radii of Fe and Cr atoms and also due to fact that both Fe and Cr atoms have identical surroundings in the unit cell. The fitting of the Cr K edge data  has been carried out using the same model as in the case  of Fe K-edge data by putting Cr as the absorbing atom at the same position of  Fe (0, 0, 0) and replacing the scattering paths by the Cr atoms according to their occupancies. In the above fitting process, no multiple scattering paths have been used. In the $\chi(R)$ versus R plots the main intense peak appears due to scattering from the 8 Co atoms at 2.44 $\mathrm{\AA}$ in the $1^{st}$ co-ordination shell. Similar to the Fe spectra, the comparatively lower intense peaks at higher have contributions of  scattering from the next two coordination shells consisting of 6 Si and 12 Fe/Cr atoms at nominal distances of 2.823 $\mathrm{\AA}$ and 3.993 $\mathrm{\AA}$ respectively. The data have been fitted without any antisite disorder and the best fit results have been summarised in the Table-III, from which it can be seen that the Co-Cr bond length also does not change much upon Cr doping and all other results are consistent with the results explained in the case of Fe K edge.\\

The most interesting result of the EXAFS study is that the degree of anti-site disorder is related to the Cr concentration. The measure of the Fe(Cr)-Si antisite disorder is related to the parameter $\delta$, which increases almost linearly with the increase in Cr concentration. As such, EXAFS helps to detect an extremely important intrinsic disorder namely "anti-site disorder" between Cr and Si sites which was undetectable by XRD and M{$\ddot{o}$}ssbauer spectroscopy techniques as reported earlier. \cite{doi:10.1063/1.2769175}

	\subsection{Magnetic Properties}
		Figure 6 shows the isothermal magnetisation curves for the $\mathrm{Co_2Fe_{1-x}Cr_xSi}$ alloys at 5 K and 300 K. A negligibly small hysteresis is seen for all x (0.1, 0.3, 0.5) values. Similar behavior is also reported in other Heusler alloys. \cite{:/content/aip/journal/apr2/3/3/10.1063/1.4959093, PhysRevB.83.184428} The saturation magnetisation values ($M_s$) at 5 K are found to be 5.5 $\mu_B/f.u.$, 4.9 $\mu_B/f.u.$  and 4.4 $\mu_B/f.u.$ for x = 0.1, x = 0.3 and x = 0.5 respectively These values are less than those predicted by the Slater-Pauling rule(5.8, 5.4 and 5.0 $\mu_B/f.u.$). The difference may be due to disorder present in these alloys.

		\begin{table*}
	\centering
	\caption{Fitting parameters for zero field electrical resistivity vs. $T$ for  $\mathrm{Co_2Fe_{0.9}Cr_{0.1}Si}$ alloy. Residual resistivity ($\rho_0$) in both cases is the same, i.e., $\rho_0$ = 37.42 $\mu\Omega$ cm}.

	\begin{tabular}{c| c| c c }
		\hline \hline& & \\
		Temperure Region & Fitting Equation & \multicolumn{2}{c}{{Fitting Parameters}} \\
		\hline & & \\
		$50 < T < 100 K$ & $\rho(T) = \rho_{0}+ AT^2 + BT^{9/2}$ &
		$ A = [2.8(6)]\times{10^{-5}}$$\mu\Omega \mathrm{\:cm\:K^{-2}}$ & $B=[8.7(8)]\times{10^{-10}}$ $\mu\Omega\:\mathrm{cm\:K^{-4.5}}$\\
		\hline  & &\\
		$ 100 K < T < 300 K$ & $\rho(T) = \rho_{0} + CT^n $ & $C= [2.88(5)]\times{10^{-4}}$ $\mu\Omega \mathrm{\:cm\:K^{-n}}$ & $ n=1.93(1)$ \\
		\hline \hline
	\end{tabular} 
	\label{} 
\end{table*}

\begin{table*}
	\centering
	\caption{Fitting parameters for zero field electrical resistivity vs. $T$ for  $\mathrm{Co_2Fe_{0.7}Cr_{0.3}Si}$ alloy. Residual resistivity ($\rho_0$) in both cases is the same, i.e., $\rho_{01}$ = 107.048 $\mu\Omega$ cm.}
	
	\begin{tabular}{c|c|c c}
		\hline\hline & & \\
		Temperure Region & Fitting Equation & \multicolumn{2}{c}{{Fitting Parameters}}\\
		\hline & & \\
		$50 < T < 100 K$ & $\rho(T) = \rho_{01}+ A_1T^2 + B_1T^{9/2}$ &
		$ A_1 = [7.4(6)]\times{10^{-5}}$$\mu\Omega \mathrm{\:cm\:K^{-2}}$ & $B_1=[4.6(8)]\times{10^{-10}}$ $\mu\Omega \mathrm{\:cm\:K^{-4.5}}$\\
		\hline  & &\\
		$ 100 K < T < 300 K$ & $\rho(T) = \rho_{01} + C_1T^n $ & $C_1= [3.6(3)]\times{10^{-5}}$ $\mu\Omega \mathrm{\:cm\:K^{-n}}$ & $ n=2.26(1)$ \\
		\hline\hline
	\end{tabular} 
	\label{} 
\end{table*}

\begin{table*}
	\centering
	\caption{Fitting parameters for zero field electrical resistivity vs. $T$ for  $\mathrm{Co_2Fe_{0.5}Cr_{0.5}Si}$ alloy. Residual resistivity ($\rho_0$) for region I and region II are found to be  75.72 and 73.85 $\mu\Omega$ cm respectively.} 
	
	\begin{tabular}{c|c|c c}
		\hline\hline & & \\
		Temperure Region & Fitting Equation & \multicolumn{2}{c}{{Fitting Parameters}}\\
		\hline & & \\
		$5 < T < 85 K$ & $\rho(T) = \rho_{03} -A_2T^{1/2}+B_2T^{9/2}$ &
		$ A_2 = [0.22(1)]$$\mu\Omega \mathrm{\:cm\:K^{-1/2}}$ & $B_2=[7.9(1)]\times{10^{-10}}$ $\mu\Omega \mathrm{\:cm\:K^{-4.5}}$\\
		\hline  & &\\
		$ 85 K < T < 300 K$ & $\rho(T) = \rho_{02} + C_2T^n $ & $C_2= [1.27(9)]\times{10^{-6}}$ $\mu\Omega \mathrm{\:cm\:K^{-n}}$ & $ n=2.75(1)$ \\		\hline\hline
	\end{tabular} 
	\label{tab5} 
\end{table*}

\subsection{Transport properties}	
Figure 8(a) shows the variation of normalised electrical resistivity (in zero magnetic field) as a function of temperature for all samples. The RRR ($\rho_{300K}/\rho_{5K}$) for $\mathrm{Co_2Fe_{1-x}Cr_xSi}$ is found to be 1.42, 1.13 and 1.09 for x = 0.1, 0.3 and 0.5 respectively. A high RRR value usually corresponds to good ordering, while a low value can be attributed to scattering from impurities or anti-site defects. In the case of Heusler alloys, the highest RRR of 6.5 is reported for $\mathrm{Co_2MnSi}$ single crystals.\cite{doi:10.1063/1.1428625} In the present alloys, a much reduced RRR may result from the presence of atomic disorder. Also, RRR value decreases with increase in Cr concentration which suggests that the disorder increases with increase in Cr concentration, as also observed from EXAFS measurements.\\

The resistivity behaviour in the temperature region from 5 - 300 K was investigated in terms of different mechanisms. Heusler alloys typically exhibit two-magnon $\mathrm{T^{9/2}}$ dependence at low temperatures and $\mathrm{T^{7/2}}$ dependence at high temperatures), electron-electron or electron-magnon ($\mathrm{T^2}$ dependence) and electron-phonon (T dependence) scattering. But, for half-metals, single magnon scattering is not possible due to the presence of the gap in the density of states in the minority sub-band.\cite{1989JPCM1.2351O,Kubo-1972,1989JPCM1.2351O} Thus, in the case of true half-metals,  $\mathrm{T^2}$ term due to single magnon scattering is expected to be absent in resistivity. The present resistivity data was analysed by fitting with different equations in different temperature regions.
\subsubsection{x=0.1 and x = 0.3}
Figures 8(b) and 8(c) show the measured and fitted temperature dependence of electrical resistivity for $\mathrm{Co_2Fe_{0.9}Cr_{0.1}Si}$ and $\mathrm{Co_2Fe_{0.7}Cr_{0.3}Si}$ respectively in the absence of magnetic field. Robustness of electrical resistivity of these alloys in magnetic fields is noticeable from the insignificant changes in $\rho(T)$ in 50 kOe, as shown in the inset of Fig. 8(b). At low temperatures ($5 K < T < 50 K$) the electrical resistivity is found to be almost independent of temperature. Similar behaviour was also reported for other half-metallic ferromagnetic Heusler alloys. \cite{PhysRevB.96.184404,PhysRevLett.110.066601} To closely investigate the half-metallic nature of this alloy, the resistivity data is fitted using suitable equations in two different temperature regimes. In the region I ($50 K < T <100 K$), the resistivity data was found to fit well with the equation
\begin{equation}
	\rho(T) = \rho_{0}+ AT^2 + BT^{9/2}
\end{equation} 
where A and B are free fitting parameters, $\rho_{0}$ is the residual resistivity arising from the scattering of conduction electrons by the lattice defects, impurities etc. Here, the $T^{9/2}$ term corresponds to two-magnon scattering and $T^2$ term corresponds to one-magnon or electron-electron scattering.
The obtained fitting parameters for x = 0.1 and x = 0.3 are given in Table IV and V respectively. The values in parenthesis specify the errors in the fitting parameters.
Generally, electron-electron (e-e) and electron-(single) magnon (e-m) contribute towards a quadratic temperature dependence to resistivity. However, between these two, the more dominant contribution can be inferred from the magnitude of $A$ ($T^2$ coefficient). A is of the order of 10$^{-2}$ n$\Omega$-cmK$^{-2}$ for e-e interaction while it is normally two orders higher for e-m interaction ($\sim n$$\Omega$-cmK$^{-2}$).\cite{Szajek2004}
As seen from table IV and V, the value of A falls within the accepted range for electron-electron scattering, indicating the absence of single magnon scattering mechanism, as expected for half-metals. The slight increase of A from x = 0.1 to 0.3 suggests the enhanced electron-electron interaction.
The resistivity data in the region II ($100 K < T < 300 K$) is found to fit well with the power law
\begin{equation}
	\rho(T) = \rho_{0}+ \rho(T) = \rho_{0} + CT^n
\end{equation}
where C is the arbitrary constant. The value of n, when fitted in the temperature range of 100-300 K is found to be 1.92 (close to 2) and 2.26 for x = 0.3 respectively. Previously, several authors have obtained different power (n) values depending on the temperature range considered. \cite{doi:10.1063/1.1739293, 0022-3727-37-15-001,doi:10.1063/1.126606}. Thus, almost $\mathrm{T^2}$ variation of resistivity in this regime suggests that the alloy with x = 0.1 retain its half-metallic character only up to 100 K. For x = 0.3, the value of n above 2 gives some signature of half-metallic character even at higher temperatures.

\subsubsection{x=0.5}
Figure 8(d) shows the measured temperature dependence of electrical resistivity for $\mathrm{Co_2Fe_{0.5}Cr_{0.5}Si}$ alloy. Resistivity decreases with temperature followed by a clear upturn below 85 K. Though such a trend can be feebly seen in the other two samples, the minimum is well defined in this case. Furthermore, application of a magnetic field is found to have no effect on the position or the depth of the minimum. This feature observed in some other bulk Heusler alloys\cite{doi:10.1063/1.4862966, doi:10.1063/1.4902831} is typically attributed to the disorder augmented coherent backscattering of conduction electrons, a mechanism known as weak localization.\cite{RevModPhys.57.287} It may be recalled that the RRR value is the lowest in x=0.5, indicating that the disorder is maximum in this composition. Though many Heusler alloy thin films have shown such minima, \cite{0022-3727-45-47-475001, doi:10.1063/1.4913516, SF} its occurrence in bulk alloys is not very common. The weak localization gives rise to $\mathrm{T^{1/2}}$ term in resistivity. Below the resisitivity minimum, i.e., for $5 K < T < 85 K$, the data fits well with equation
\begin{equation}
	\rho(T) = \rho_0+ \rho(T) = \rho_{03} -A_2T^{1/2}+B_2T^{9/2}
\end{equation}

In the temperature region $85 K < T < 300 K$, the resistivity data is found to fit with the power law with n = 2.75.
The evaluated fitting parameters are given in Table VI. In the region I, the dominant contribution is from the $\mathrm{T^{1/2}}$ term which is due to disorder. In the region II, the $\rho$ vs. T curve follows a power law with n = 2.75. This power law is not associated with any specific scattering mechanism but it can represent a combination of electron-electron, electron-phonon and electron-magnon scatterings. Similar values of n are also reported for $\mathrm{Co_{2-x}Fe_{1+x}Si}$ alloys. \cite{doi:10.1063/1.4813519} Thus, for x = 0.5 , the value of n above $2$ gives signature of half metallic character even at high temperatures similar to x = 0.3.
It is interesting to see that even though the strength of disorder increases with Cr substitution, the alloys retain their half-metallic characteristics. Thus, Cr substitution provides a much robust half-metallicity making these alloys useful for spintronic applications.

\subsection{Theoretical Results}
	The effect of Cr substitution do not seem to be significant on the structural parameters. Computationally  optimized lattice parameters are found to almost remain unchanged (ranging from $5.62$\r{A} to $5.63$ \r{A}) as we go from $x=0$ to $x=1$ in Co$_2$Fe$_{1-x}$Cr$_x$Si, which is in good agreement with the previously reported data.\cite{guezlane2016electronic, 6PhysRevB.72.184434} Co$_2$MSi, (M= Fe, Mn, etc.) are very well studied compounds, being potential candidates for various spintronic applications. They usually show metallic behaviour in majority spin channel while giving significant band gap in minority spin channel.  Various theoretical studies\cite{guezlane2016electronic,doi:10.1063/1.2769175} show that the calculated electronic structure for these systems using PBE functional can lead to incorrect predictions about the half metallicity. An improvement on PBE is a GGA+U\cite{doi:10.1063/1.2769175} calculations, which better describes the magnetic and electronic properties. GGA+U, however, involves a Hubbard 'U' parameter, which is a strongly system dependent parameter and is freely tunable to fit the experimentally measured values. HSE-06 functional, on the other hand, usually predicts better band gap value for experimentally unknown compounds.\cite{shi2017hybrid, gao2016monolayer,morales2017empirical} That is why, to gain a more accurate insight on the effect of Cr substitutions in Co$_2$FeSi, we have simulated the spin-polarized electronic structure of Co$_2$Fe$_{1-x}$Cr$_x$Si, (x = 0, 0.125, 0.25, 0.50 and  1) using HSE-06 functional. Figure 9 shows the density of states (dos) for majority and minority spin channels for the above compounds. 
	
	We also compared the total magnetic moment per formula unit, and minority band gap for all the alloy concentrations using the PBE-GGA and HSE-06 functional, as tabulated in Table VII. Our calculated magnetic data matches almost exactly with previously reported data using PBE-GGA function. While compared to Slater-Pauling estimated values, one can see that HSE-06 slightly overestimates the magnetic moment for all $x$. Interestingly, with increasing Cr concentrations, the overestimation of total moment by HSE-06 functional is larger. Compared to experimental data (see Fig. 7 for moments at x = 0.1,0.3 0.5), however, simulated total moments from both the functionals turn out to be larger. This discrepancy can be attributed to the increase of small disorder with increase in Cr concentration in prepared sample, as compared to the completely ordered sample used to simulate the theoretical data. Having a closer  look at the orbital contribution to band edges reveals that the valence band maxima (VBM) mainly consists of Co-d orbitals, while the conduction band minima (CBM) has contribution from both Co-d and Fe-d (Cr-d) for pure Co$_2$FeSi (Co$_2$CrSi). In case of Co$_2$Fe$_{1-x}$Cr$_x$Si alloys, CBM has negligible contribution from Cr-d, and is mainly comprised of Co-d and Fe-d orbitals. By looking at TABLE 1, one can see that calculated minority band gaps using HSE-06 are much higher compared to PBE ones. The most striking difference between the PBE and HSE-06 results  is the emergence of a small but finite state at the Fermi level (E$_F$) in the dos of minority spin channel in case of PBE-GGA for x = 0 to 0.25, \cite{PhysRevLett.110.066601} unlike the HSE-06 case. This actually predicts non half-metallic behavior, unlike the experimental observation. Similar scenario arises in case of using metaGGA (MBJ) functional for these alloys.\cite{guezlane2016electronic} In case of HSE-06 (see Fig. 9), however, it can be clearly seen tha all pure systems and alloys have significant minority band gap, agreeing well with the experiment\cite{PhysRevLett.110.066601} From the predicted value of the minority gap, these compounds can be seen as excellent candidates for spintronic applications.

\begin{table}[t!]
\caption{Calculated data for total magnetic moment ($\mu_B$/ formula unit) and band gap in minority spin channel (E$_g$$^\downarrow$) for Co$_2$Fe$_{1-x}$Cr$_x$Si, (x=0, 0.125, 0.25, 0.50 and  1) using both PBE and HSE-06 exchange correlation functional. }	
	\begin{ruledtabular}
		\begin{centering}
			\begin{tabular}{c c c  c c}
				\multirow{2}{*}{x}
				& \multicolumn{2}{c}{PBE} & \multicolumn{2}{c}{HSE-06}  \\
				
				\vspace{0.08 in}
				&  ($\mu_B$/f.u.) & E$_g$$^\downarrow$ (eV) & ($\mu_B$/f.u.) & E$_g$$^\downarrow$ (eV) \\
				\hline
				\vspace{0.08 in} 
				0 & 5.48 & 0.09 & 6.20 & 1.82  \\
				\vspace{0.08 in}
				0.125 & 5.41 & 0.05 & 5.97 & 1.77 \\
				\vspace{0.08 in}
				0.25 & 5.33 & 0.03 & 5.73 & 1.74 \\
				\vspace{0.08 in}
				0.50 & 4.98 & 0.46 & 5.24 & 2.01 \\
				\vspace{0.08 in}
				1 &  4.00 & 0.72 & 4.24 & 2.66 \\
			\end{tabular}
		\end{centering}
		
	\end{ruledtabular}
\end{table}
		
\section{Conclusion}
		In conclusion, we have studied the effect of Cr substitution for Fe on the structural, magnetic and transport properties of $\mathrm{Co_2FeSi}$ Heusler alloy. The alloys crystallize in the cubic structure for all values of $x$. The EXAFS results reveal that the anti-site disorder increases almost linearly with inceasing Cr concentration. The isothermal magnetization curves show that the saturation magnetization decreases with Cr concentration and the values at 5 K are found to be low in comparison to those predicted by the Slater-Pauling rule. The deviation from the Slater-Pauling rule is found to increases with the increase of Cr concentration. Such deviation is attributed to the enhancement of small disorder in the prepared sample.The resistivity measurements reveal that for x = 0.1, the half-metallic character is retained only up to 100 K, but for x = 0.3 and 0.5, it is seen to be present even at higher temperatures. The residual resistivity value is found to decrease with Cr substitution which also suggests the enhancement of disorder with increasing Cr concentration. The minima in the $\rho$ - T plot for x = 0.5 is attributed to the weak localization resulting from the high degree of disorder. Even though the disorder increases with Cr substitution, the half-metallic character still persists and thus, the half-metallicity in these alloys is quite robust against anti-site disorder. First principles calculations based on the hybrid exchange correlation functional (HSE-06)  reveals the shortcomings of the previously reported predictions done by usual local density or generalized gradient approximation. Earlier studies show the emergence of small finite states at E$_F$ in the minority spin channel making the system shy from being half metallic. Use of HSE-06 functional, however, removes this error and predicted a large (zero DoS) finite band gap in the minority channel, confirming the half metallic behavior for all the compounds, as observed experimentally. Calculated band gap and magnetic moments for all the alloys are also compared with those measured, wherever applicable. All these properties, with a confirmation from both theory and experiment, make $\mathrm{Co_2Fe_{1-x}Cr_xSi}$ a promising material for spintronics application.\\

\section{Acknowledgments}
DR would like to thank Council of Scientific and Industrial Research (CSIR), India for providing Junior Research Fellowship and Dr. S. Shanmukharao Samatham for helping in resistivity measurement.

\bibliography{bib}

\end{document}